\documentclass{PoS}
\usepackage{amsmath}
\usepackage{amssymb}
\usepackage{booktabs}

\title{Matrix elements of the electromagnetic operator between kaon and pion states}

\ShortTitle{Electromagnetic operator between kaon and pion states}

\author{\speaker{I. Baum}, G.~Martinelli\\
Dipartimento di Fisica, Universit{\`a} di Roma ``La Sapienza'' and INFN - Sezione di Roma, P.le A. Moro 5, I-00185 Roma, Italy}

\author{V.~Lubicz\\
Dipartimento di Fisica, Universit{\`a} di Roma Tre and INFN - Sezione di Roma Tre, Via della Vasca Navale 84, I-00146 Roma, Italy}

\author{S.~Simula\\
INFN, Sezione di Roma Tre, Via della Vasca Navale 84, I-00146 Roma, Italy}

\author{for the ETM Collaboration}

\abstract{We compute the matrix elements of the electromagnetic (EM) operator between kaon and pion states, using 
lattice QCD with maximally twisted-mass fermions and two flavors of dynamical quarks ($N_f = 2$). 
The EM operator is renormalized non-perturbatively in the RI'/MOM scheme and our simulations cover pion masses 
as light as 260 MeV and three values of the lattice spacing, ranging from $\sim$~0.07 up to $\sim$~0.1 fm.
At the physical point our preliminary result for the $K \to \pi$ tensor form factor at zero-momentum transfer is $f_T^{K\pi}(0) 
= 0.42(2_{\rm stat})$, which differs significantly from the old quenched result $f_T^{K\pi}(0) = 0.78(6)$ obtained by the 
SPQcdR Collaboration \cite{SPQCDR} with pion masses above 500 MeV. 
We investigate the source of this difference and conclude that it is mainly related to the chiral extrapolation of the quenched 
data.
For the case of the tensor charge of the pion we obtain the preliminary value $f_T^{\pi\pi}(0) = 0.200(14_{\rm stat})$, which 
can be compared with the result $f_T^{\pi\pi}(0) = 0.216(34)$ obtained at $N_f = 2$ by the QCDSF Collaboration \cite{QCDSF} 
using higher pion masses.}

\FullConference{The XXVIII International Symposium on Lattice Field Theory, Lattice2010\\
		June 14-19, 2010\\
		Villasimius, Italy}

\begin{document}

\section{Introduction}
Precision measurements of weak decays can constrain the parameters of the Standard Model (SM) and place
bounds on new physics (NP) models, such as supersymmetry. In particular, penguin operators between
kaon and pion states can place strong bounds on the CP-violating parameters in the light quark sector.

In this contribution we present a lattice study of the electromagnetic (EM) operator, relevant in the CP violating 
part of the $K \to \pi \ell^+ \ell-$ semileptonic decays, performed using the gauge configurations generated by 
the European Twisted Mass Collaboration (ETMC) with $N_f = 2$ maximally twisted-mass fermions.
The EM operator involved in the weak $s \to d$ transition is given by
 \begin{eqnarray}
     Q_{EM} = \bar{s} ~ F_{\mu \nu} \sigma^{\mu \nu} d ~ ,
     \label{eq:EM}
 \end{eqnarray}
where $F_{\mu \nu}$ is the EM field tensor. Therefore its matrix element between kaon and pion states 
involves the one of the weak tensor current, which can be written in terms of a single form factor, 
$f_T^{K\pi}(q^2)$, as
 \begin{eqnarray}
      \langle \pi^0(p_{\pi}) | \bar{s} \sigma^{\mu \nu} d | K^0(p_{K}) \rangle = (p_{\pi}^\mu p_{K}^\nu - 
      p_{\pi}^\nu p_{K}^\mu) ~ \frac{ \sqrt{2} f_T^{K\pi}(q^2)}{M_K + M_\pi} ~  ,
      \label{eq:EM_KPi}
 \end{eqnarray}
where $q^\mu \equiv (p_K - p_\pi)^\mu$ is the 4-momentum transfer. 
Note that the mass factor $(M_K + M_\pi)^{-1}$ is conventionally inserted in Eq.~(\ref{eq:EM_KPi}) 
in order to make the tensor form factor dimensionless.

Our  simulations cover pion masses as light as 260 MeV and three values of the lattice spacing, ranging 
from $\sim$~0.07 up to $\sim$~0.1 fm.
At the physical point  our preliminary result for the $K \to \pi$ tensor form factor at zero-momentum 
transfer is 
 \begin{eqnarray}
     f_T^{K\pi}(0) = 0.42 ~ (2_{\rm stat}) \qquad \qquad \mbox{(ETMC)} ~ ,
     \label{eq:fTKPi}
 \end{eqnarray}
 where the error is statistical only. 
 Our finding (\ref{eq:fTKPi}) differs significantly from the old quenched result $f_T^{K\pi}(0) = 0.78(6)$ obtained 
 in Ref.~\cite{SPQCDR} by the SPQcdR Collaboration with pion masses above $\sim$~500 MeV. 
The reason is mainly due to the non-analytic behavior of the tensor form factor $f_T^{K\pi}(0)$ in terms of the quark 
masses introduced by the mass factor $(M_K + M_\pi)^{-1}$ in the parameterization (\ref{eq:EM_KPi}). 
Such a behavior was not taken into account in Ref.~\cite{SPQCDR} (see later on).

In the case of the degenerate $\pi \to \pi$ transition, making use of the pion mass dependence predicted by  Chiral 
Perturbation Theory (ChPT) in Ref.~\cite{SU2_ChPT}, we obtain for the tensor form factor $f_T^{\pi\pi}(0)$, 
also known as the tensor charge of the pion, the preliminary value at the physical point
  \begin{eqnarray}
     f_T^{\pi\pi}(0) = 0.200 ~ (14_{\rm stat})  \qquad \qquad \mbox{(ETMC)} ~ ,
     \label{eq:fTPiPi}
 \end{eqnarray}
 which can be compared with the result $f_T^{\pi\pi}(0) = 0.216(34)$ obtained at $N_f = 2$ by the QCDSF Collaboration 
 \cite{QCDSF} using simulations at higher pion masses.

\section{$K \to \pi$ results}

We have performed the calculations of all the relevant 2-point and 3-point correlation functions using the ETMC gauge 
configurations with $N_f = 2$ dynamical twisted-mass quarks \cite{ETMC_scaling} generated at three values of $\beta$, 
namely the ensembles $A_2 - A_4$ at $\beta = 3.8$ ($a \simeq 0.103$ fm), $B_1 - B_7$ at $\beta = 3.9$ ($a \simeq 
0.088$ fm), and $C_1 - C_3$ at $\beta = 4.05$ ($a \simeq 0.070$ fm). 
The pion mass $M_\pi$ ranges between $\simeq 260$ MeV and $\simeq 575$ MeV and the size $L$ of our lattices 
guarantees that  $M_\pi L$ is larger than $\sim 3.3$. 
For each pion mass and lattice spacing we have used several values of the (bare) strange quark mass $m_s$ to allow 
for a smooth, local interpolation of our results to the physical value of $m_s$ (see Ref.~\cite{ETMC_masses}).
The calculation of the 2- and 3-point correlation functions has been carried out using all-to-all quark propagators evaluated 
with the {\it one-end-trick} stochastic procedure and adopting non-periodic boundary conditions which make arbitrarily small 
momenta accessible. 
All the necessary formulae can be easily inferred from Ref.~\cite{ETMC_pion}, where the degenerate case of the vector pion 
form factor is illustrated in details. 
For each pion mass the statistical errors are evaluated with the jackknife procedure.

The tensor current was renormalized non-perturbatively in the RI'/MOM scheme in Ref.~\cite{EMO_renorm}, 
including ${\cal{O}}(a^2)$ corrections coming from lattice perturbation theory \cite{EMO_perturb}.
The numerical values used in our analyses for the tensor renormalization constant are $Z_T(\rm{\overline{MS}, 2 ~ GeV}) = 
0.733(9)$, $0.743(5)$, $0.777(6)$ for $\beta = 3.8, ~ 3.9, ~ 4.05$, respectively.

\begin{figure}[!htb]
\centerline{\includegraphics[width=14cm]{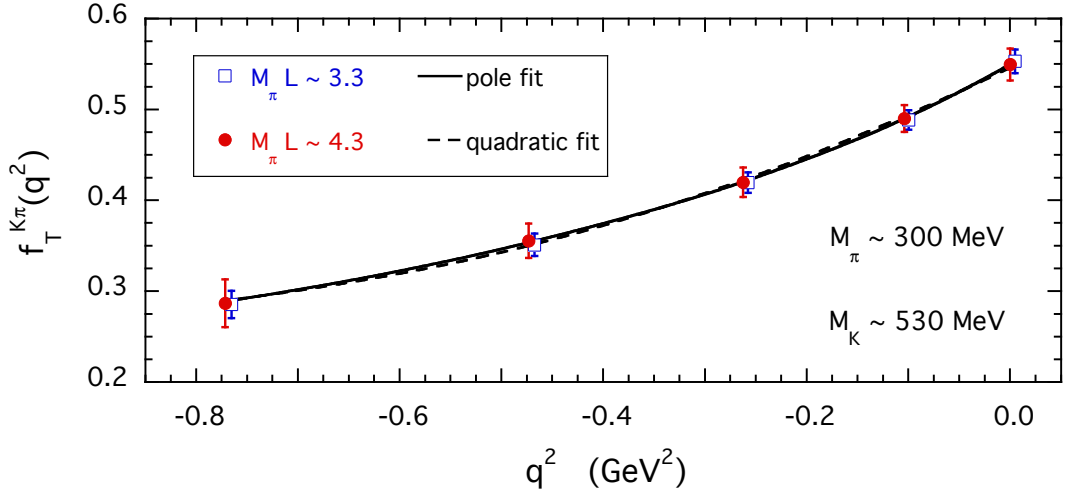}}
\caption{\label{fig:fT_q2} The tensor form factor $f_T^{K\pi}(q^2)$ obtained at $M_\pi \simeq 300$ MeV and 
$M_K \simeq 530$ MeV versus $q^2$ in physical units.
The dots and the squares (shifted for better clarity) correspond to the gauge ensembles $B_1$ and $B_7$, 
respectively, which differs only for the lattice size.
The solid and dashed lines are the results of the fits based on Eqs.~(\protect\ref{eq:pole}) and 
(\protect\ref{eq:quadratic}), respectively.}
\end{figure}

At each pion and kaon masses we determine the tensor form factor $f_T^{K\pi}(q^2)$ for several values of 
$q^2 < q_{max}^2 = (M_K - M_\pi)^2$ in order to interpolate at $q^2 = 0$. 
Note that, because of the vanishing of the Lorentz structure in Eq.~(\ref{eq:EM_KPi}), it is not possible 
to determine $f_T^{K\pi}(q^2)$ at $q^2 = q_{max}^2$.
In this respect we take advantage of the non-periodic boundary conditions to reach values of $q^2$ quite 
close to $q^2 = 0$.
The momentum dependence of $f_T^{K\pi}(q^2)$ can be nicely fitted either by a 
pole behavior 
   \begin{eqnarray}
    f_T^{K\pi}(q^2) = f_T^{K\pi}(0) / (1 - s_T^{K\pi} ~ q^2)
    \label{eq:pole}
   \end{eqnarray}
or by a quadratic fit in $q^2$
  \begin{eqnarray}
    f_T^{K\pi}(q^2) = f_T^{K\pi}(0) \cdot (1 + s_T^{K\pi} ~ q^2 + c_T^{K\pi} ~ q^4) ~ .
    \label{eq:quadratic}
   \end{eqnarray}
The good quality of both fits is illustrated in Fig.~\ref{fig:fT_q2}, where the results obtained at two different 
lattice volumes are also compared.
It can clearly be seen that: ~ i) finite size effects are well below the statistical precision of our lattice points;
~ ii) the results for $f_T^{K\pi}(0)$ (as well as those for the slope $s_T^{K\pi}$), obtained using the pole 
dominance (\ref{eq:pole}), differ only very slightly from those obtained via the quadratic fit (\ref{eq:quadratic}).
Therefore in what follows we limit ourselves to the results for both $f_T^{K\pi}(0)$ and the slope $s_T^{K\pi}$ 
obtained through the pole fit (\ref{eq:pole})

\begin{figure}[!htb]
\centerline{\includegraphics[width=14cm]{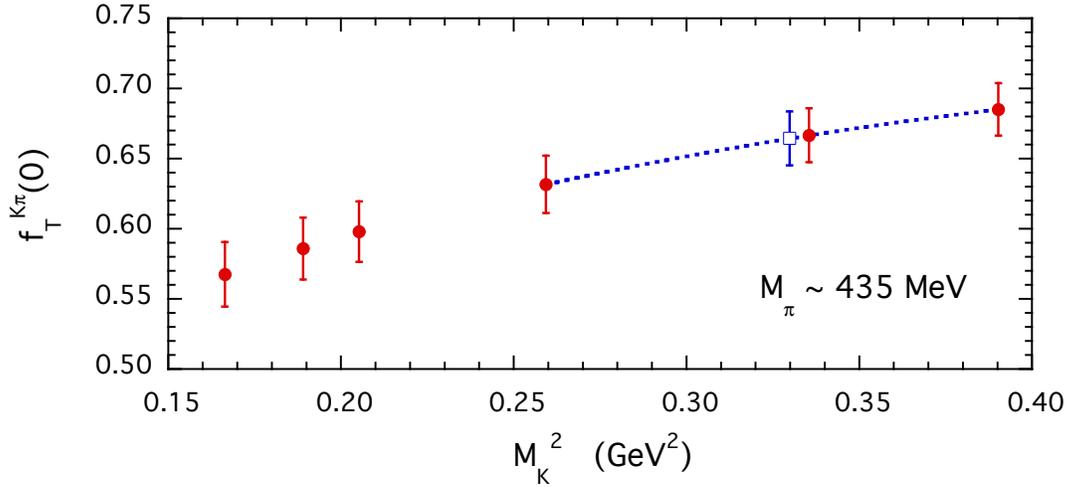}}
\caption{\label{fig:fT0} Results for $f_T^{K\pi}(0)$ versus $M_K^2$ at $M_\pi \simeq 435$ MeV.
The square corresponds to the value of $f_T^{K\pi}(0)$ obtained by local interpolation via 
quadratic splines (dotted line) at the reference kaon mass $M_K^{ref} \simeq 575$ 
MeV from Eq.~(\protect\ref{eq:MKref}).}
\end{figure}

The values obtained for $f_T^{K\pi}(0)$ and $s_T^{K\pi}$ depend on both pion and kaon masses.
The dependence on the latter is shown in Fig.~\ref{fig:fT0} for $f_T^{K\pi}(0)$ at $M_\pi \simeq 435$ MeV 
and it appears to be quite smooth. 
Thus an interpolation at the physical strange quark mass can be easily performed using quadratic splines.
This is obtained by fixing the combination ($2 M_K^2 - M_\pi^2$) at its physical value, which at each pion 
mass defines a {\it reference} kaon mass, $M_K^{ref}$, given by
 \begin{eqnarray} 
     2 [M_K^{ref}]^2 - M_\pi^2 = 2 [M_K^{phys}]^2 - [M_\pi^{phys}]^2 
     \label{eq:MKref}
 \end{eqnarray}
with $M_\pi^{phys} = 135.0$ MeV and $M_K^{phys} = 494.4$ MeV.

\begin{figure}[!htb]
\centerline{\includegraphics[width=14cm]{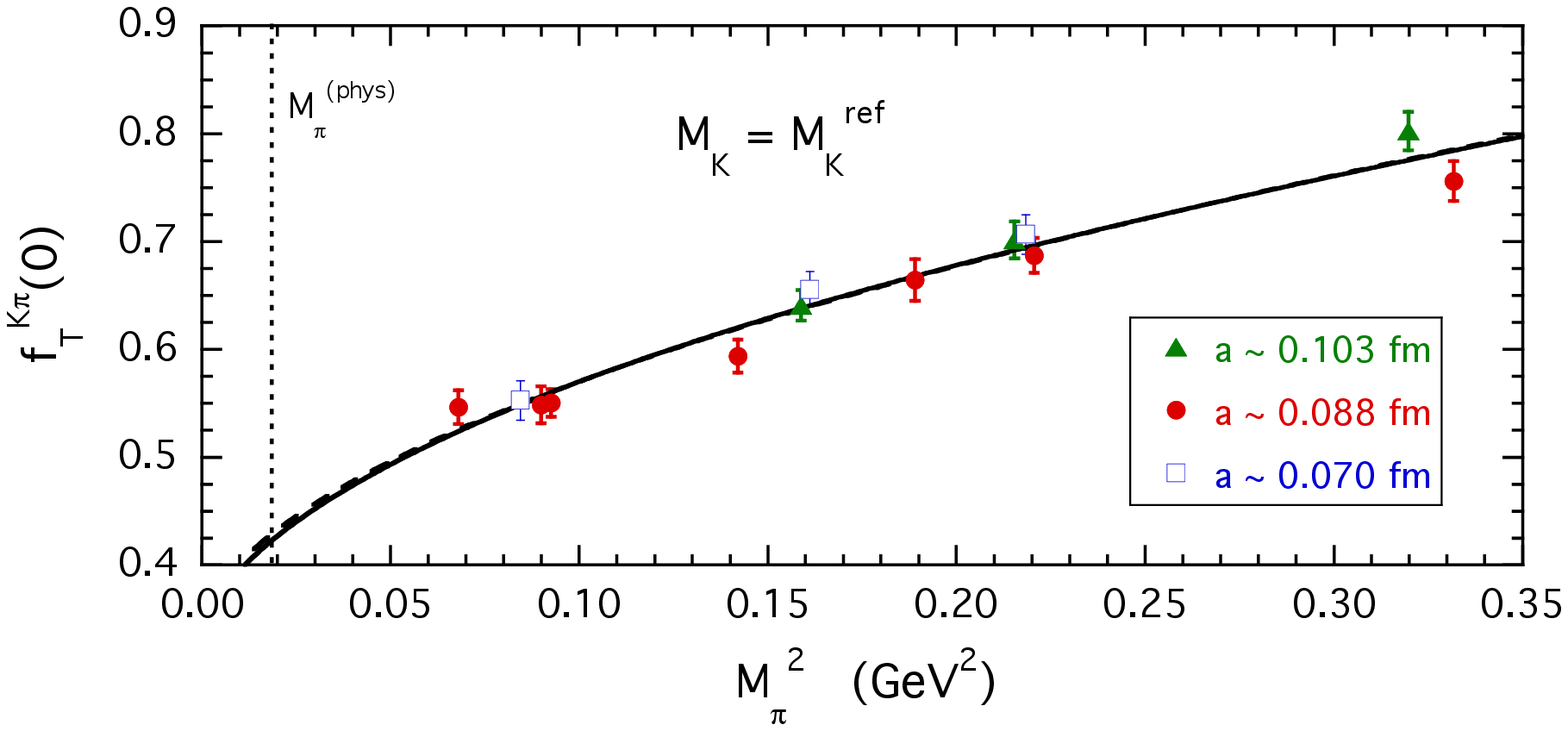}}
\caption{\label{fig:fT(0)} Results for $f_T^{K\pi}(0)$ versus $M_\pi^2$ at $M_K = M_K^{ref}$ in 
physical units.
The dots, squares and triangles are our results for the three lattice spacings of the ETMC 
simulations, specified in the inset.
The solid and dashed lines correspond to the fit given by Eq.~(\protect\ref{eq:chiral_fit}) with 
$B = 0$ and $D = 0$, respectively.}
\end{figure}

\begin{figure}[!htb]
\centerline{\includegraphics[width=14cm]{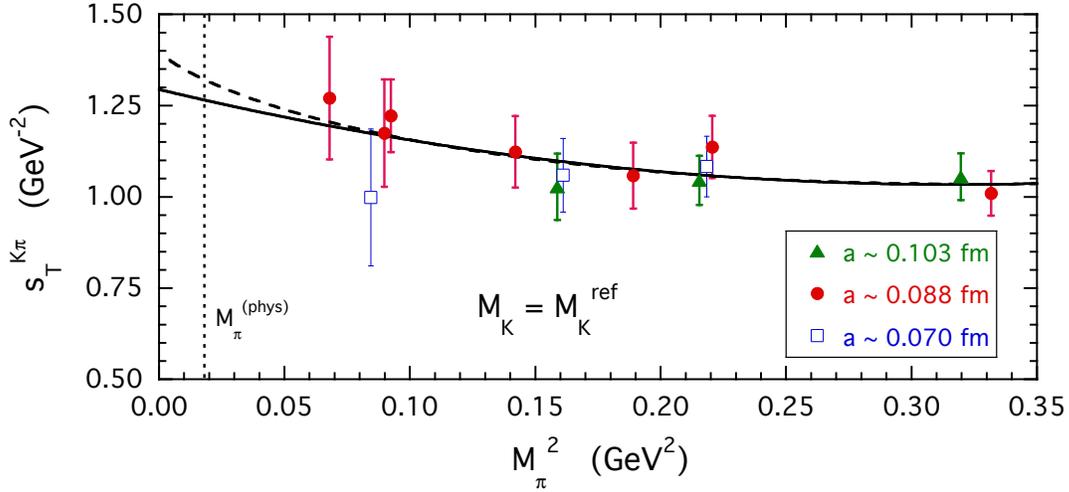}}
\caption{\label{fig:sT} The same as in Fig.~\protect\ref{fig:fT(0)}, but for the slope $s_T^{K\pi}$ of 
the tensor form factor at $q^2 = 0$.}
\end{figure}

The results for $f_T^{K\pi}(0)$ and the slope $s_T^{K\pi}$, interpolated at the reference kaon mass 
$M_K = M_K^{ref}$, are shown in Figs.~\ref{fig:fT(0)} and \ref{fig:sT}, respectively, for the three lattice 
spacings of our simulations.
It can clearly be seen that discretization effects are sub-dominant and therefore, in what follows, we 
concentrate on the chiral extrapolation of our lattice data, which in the case of $f_T^{K\pi}(0)$ is a 
much more delicate point.

In Ref.~\cite{SPQCDR} the first lattice calculation of the EM operator matrix element between kaon 
and pion states was carried out in the quenched approximation and for pion masses above 
$\sim$~500 MeV. There the chiral extrapolation was performed assuming that $f_T^{K\pi}(0)$ 
reaches a non-vanishing value in the SU(3) chiral limit $(M_K, ~ M_\pi) \to 0$. 
A simple linear fit in the squared kaon and pion masses was attempted obtaining at the physical 
point the result $f_T^{K\pi}(0) = 0.78(6)$.

In the degenerate case $M_K = M_\pi$ the chiral expansion of the tensor current  has been 
studied in Ref.~\cite{SU2_ChPT}. 
The main finding is that the form factor $f_T^{\pi\pi}(0)$ vanishes like $M_\pi$ for $M_\pi \to 0$, 
so that the ratio $f_T^{\pi\pi}(0) / M_\pi$ tends to a non-vanishing value in the chiral limit.
The same argument is expected to hold as well in the case of the $K \to \pi$ transition: the 
form factor $f_T^{K\pi}(0)$ must vanish in the SU(3) chiral limit in such a way that the ratio 
$f_T^{K\pi}(0) / (M_K + M_\pi)$ does not vanish in limit $M_K = M_\pi = 0$.

Therefore we perform the chiral extrapolation of our lattice data using the ansatz
 \begin{eqnarray}
     f_T^{K\pi}(0) = (M_K^{ref} + M_\pi) ~ A ~ \left[ 1 + B M_\pi^2 ~ \mbox{log}(M_\pi^2) + 
                                C M_\pi^2 + D M_\pi^4 \right] ~ ,
     \label{eq:chiral_fit}
 \end{eqnarray}
where $A$, $B$, $C$ and $D$ are unknown low-energy constants (LECs).
The results of the fit (\ref{eq:chiral_fit}) assuming either $B = 0$ (no chiral logs) or $D = 0$ are shown 
in Fig.~\ref{fig:fT(0)} by the solid and dashed lines, respectively. 
It can be seen that the effects of the chiral logs are not visible in our data.
At the physical point we get 
 \begin{eqnarray}
     f_T^{K\pi}(0) = 0.42(2_{\rm stat})  \qquad \qquad \mbox{(ETMC)} ~ ,
     \label{eq:KPi_ETMC}
 \end{eqnarray}
where the error is statistical only.
Had we neglected the mass factor ($M_K^{ref} + M_\pi$) in Eq.~(\ref{eq:chiral_fit}) the result at the 
physical point would change only marginally: $f_T^{K\pi}(0) = 0.46(2_{\rm stat})$.

On the contrary, in the case of the quenched data of Ref.~\cite{SPQCDR}, which were determined 
at pion masses above $\sim$~500 MeV, the inclusion of the mass factor ($M_K + M_\pi$) in the 
chiral extrapolation changes significantly the result at the physical point by many standard 
deviations, namely from $f_T^{K\pi}(0) = 0.78(6)$ to
 \begin{eqnarray}
      f_T^{K\pi}(0) = 0.49(4)  \qquad \qquad \mbox{(SPQcdR)} ~ .
      \label{eq:KPi_SPQCDR}
 \end{eqnarray}
These findings indicate that the real effect of the quenched approximation does not exceed 
$15 \%$, provided the correct mass factor is included in the chiral extrapolation. 

In the case of the slope $s_T^{K\pi}$ no mass factor should be considered and the chiral 
extrapolation of the lattice data shown in Fig.~\ref{fig:sT} provides at the physical point the 
value $s_T^{K\pi} = 1.29(18_{\rm stat})$ GeV$^{-2}$, which is consistent within the errors 
with the quenched result of Ref.~\cite{SPQCDR} $s_T^{K\pi} = 1.11(5)$ GeV$^{-2}$.

\section{Pion tensor charge}

Following Ref.~\cite{SU2_ChPT} the chiral expansion of the pion tensor charge $f_T^{\pi\pi}(0)$ 
has the form
 \begin{eqnarray}
     f_T^{\pi\pi}(0) = M_\pi ~ A' ~ \left[ 1 + \frac{M_\pi^2}{(4 \pi f_\pi)^2} ~ \mbox{log}(M_\pi^2) + C' M_\pi^2 + 
                                 D' M_\pi^4 \right] ~ ,
     \label{eq:chiral_fit_pion}
 \end{eqnarray}
where $f_\pi \sim 130$ MeV and the presence of the mass factor $M_\pi$ is expected to have an important, 
bending effect on the value extrapolated at the physical point.

\begin{figure}[!htb]
\centerline{\includegraphics[width=14cm]{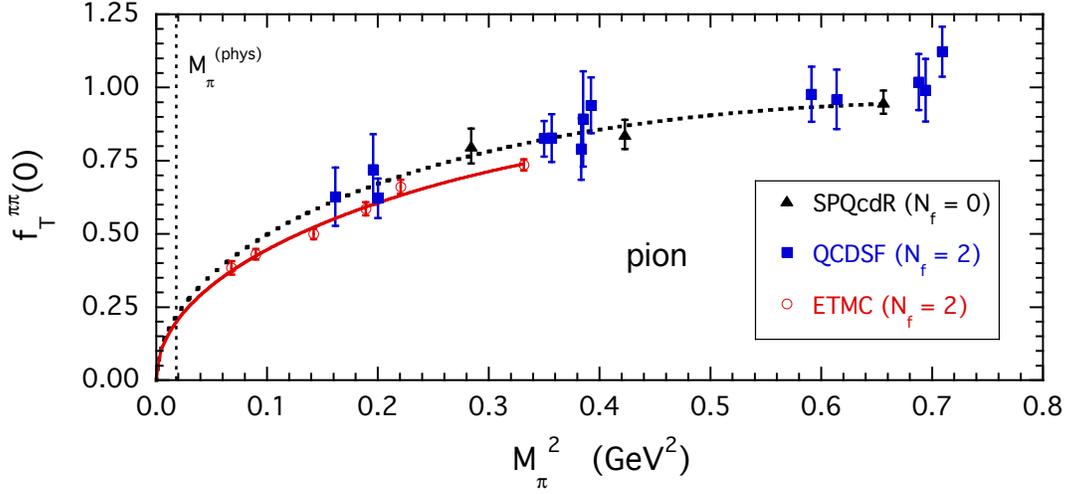}}
\caption{\label{fig:pion} Results for the pion tensor charge $f_T^{\pi\pi}(0)$ versus $M_\pi^2$ in 
physical units from our simulations at $\beta = 3.9$ (dots) and from Refs.~\protect\cite{SPQCDR} 
(triangles) and \protect\cite{QCDSF} (squares).
The solid line is the result of the fit (\protect\ref{eq:chiral_fit_pion}) applied to our lattice points, while 
the dotted line corresponds to the fit described in the text and applied to the quenched data of 
Ref.~\protect\cite{SPQCDR}.}
\end{figure}

Our results for $f_T^{\pi\pi}(0)$, obtained at $\beta = 3.9$ for $260~\rm{MeV} \lesssim M_\pi \lesssim 
575~\rm{MeV}$, are shown in Fig.~\ref{fig:pion} and compared with the ones from Ref.~\cite{QCDSF}, 
having $M_\pi \gtrsim 440$ MeV, and with the quenched calculations of Ref.~\cite{SPQCDR}, 
ranging from $M_\pi \sim 530$ MeV up to $M_\pi \sim 800$ MeV.
It can be seen that our results have a better statistical precision and cover much lighter pion masses, 
where the bending effect due to the overall mass factor $M_\pi$ is clearly visible.

Using Eq.~(\ref{eq:chiral_fit_pion}) with our lattice points, we get at the physical point 
 \begin{eqnarray}
     f_T^{\pi\pi}(0) & = & 0.200~(14_{\rm stat}) \qquad \qquad \mbox{(ETMC)} ~ , 
     \label{eq:pion_ETMC}
 \end{eqnarray}
 to be compared with the QCDSF result \cite{QCDSF} at $N_f = 2$ 
 \begin{eqnarray}
     f_T^{\pi\pi}(0)| & = & 0.216 ~ (34) \qquad \qquad \mbox{(QCDSF)} ~ . 
     \label{eq:pion_QCDSF}
 \end{eqnarray}
Finally, we also apply a simple fit of the form $f_T^{\pi\pi}(0) = M_\pi ~ A' ~ \left[ 1 + C' M_\pi^2 \right]$ 
to the three quenched data of Ref.~\cite{SPQCDR}, obtaining at the physical point the result
 \begin{eqnarray}
     f_T^{\pi\pi}(0) & = & 0.221~(21_{\rm stat}) \qquad \qquad \mbox{(SPQcdR)} ~ , 
     \label{eq:pion_SPQCDR}
 \end{eqnarray}
which clearly shows that quenching effects are sub-dominant on the pion tensor charge.

\end{document}